\documentclass[useAMS,usenatbib]{mnras}
\usepackage{epsfig,graphicx,latexsym,amsmath,amssymb}
\usepackage[export]{adjustbox}[2011/08/13]

\usepackage{natbib}
\usepackage{hyperref}
\usepackage{mathrsfs}
\usepackage{lastpage}
\usepackage{xcolor}

\usepackage{color}

\bibpunct[,]{(}{)}{;}{a}{}{,}

\title[IMRIs in merging elliptical galaxies]
      {Intermediate-mass ratio inspirals in merging elliptical galaxies} %
\author[V\'{a}zquez-Aceves et al.] 
{Ver\'{o}nica V\'{a}zquez-Aceves$^{1}$
                        \thanks{E-mail: veronica@nao.cas.ac (VVA)},
Pau Amaro Seoane$^{2,3,4,5,6}$,
Dana Kuvatova$^{6}$, 
Maxim Makukov$^{6}$, \newauthor
Chingis Omarov$^{6}$, 
Denis Yurin$^{6}$
\\
$^{1}$Institute of Applied Mathematics, Academy of Mathematics and Systems Science,
CAS, Beijing 100190, China\\
$^{2}$Institute for Multidisciplinary Mathematics, UPV, València, Spain\\
$^{3}$Lanzhou Center for Theoretical Physics, Key Laboratory of Theoretical Physics of Gansu Province, School of Physical Science and Technology,\\ Lanzhou University, Lanzhou 730000, People's Republic of China\\
$^{4}$Institute of Theoretical Physics \& Research Center of Gravitation, Lanzhou University, Lanzhou 730000, People's Republic of China\\
$^{5}$Kavli Institute for Astronomy and Astrophysics, Beijing 100871, China\\
$^{6}$Fesenkov Astrophysical Institute 050020 Almaty, Kazakhstan
}

\begin{document}

\date{draft \today}

\label{firstpage}
\pagerange{\pageref{firstpage}--\pageref{lastpage}}
\maketitle
\begin{abstract}
Close encounters between two initially unbound objects can result in a binary system if enough energy is released as gravitational waves (GWs). We address the scenario in which such encounters occur in merging elliptical galaxies. There is evidence that elliptical galaxies can harbor intermediate-mass black holes. Therefore, these systems are potentially the breeding grounds of sources of gravitational waves corresponding to inspiraling compact objects onto a massive black hole due to the dynamics, the large densities, and the number of compact remnants they contain. We show that this process is efficient for intermediate-mass black holes (IMBHs) with masses ranging from M $\in (10^3,10^5)$ M$_{\odot}$ and results in the formation of intermediate mass-ratio inspirals (IMRIs). We consider a set of IMBHs and smaller black holes with masses $m_2 \in (10,10^3)$ M$_{\odot}$ to estimate the IMRI formation rate. We find rates ranging between 10$^{-8}$ yr$^{-1}$, and 10$^{-5}$ yr$^{-1}$, and the IMRI formation rate per comoving volume in merging galaxies as a function of the redshift. The peak frequencies of the gravitational radiation emitted when these IMRIs are formed are within the detection band of space-borne detectors such as LISA and TianQin; taking into account the observable volume of these detectors, the total amount of IMRI detections per year is significant. 
\end{abstract}


\section{Introduction}
\label{sec.intro}
Elliptical galaxies can be found in many sizes; their stellar population is composed mainly of old stars, making them ideal candidates for harboring compact remnants. One of the largest galaxies known in the universe is the elliptical galaxy $M87$; its total mass is about $2\times10^{12}$ M$_{\odot}$ \citep{M87mass} and harbors the first observed supermassive black hole (SMBH) $M_{87^*}$, with a mass of $6.5\times10^9$ M$_{\odot}$ \citep{M87}. On the contrary, dwarf elliptical galaxies can have masses as low as $\sim 10^6$ M$_{\odot}$\citep{Mateo_1998}. Due to the large variety in size of elliptical galaxies and the observed correlation between a galaxy and its central massive black hole (MBH) \citep{Ferrarese_2000,McConnell_2013}, it is expected to find not only SMBH but also intermediate-mass black holes (IMBHs) with masses ranging from $10^{3}$ to $10^{5}$ M$_{\odot}$. The existence and formation channels of such IMBHs remain uncertain, but observational studies suggest their presence in several galaxies; for example, there is evidence of an IMBH of $\sim 1\times 10^5$ M$_{\odot}$ in the dwarf elliptical galaxy POX 52 \citep{Barth_2004}, and of an IMBH of $\lesssim 3.8\times 10^4$ M$_{\odot}$ in the elliptical galaxy NGC 205 \citep{Valluri_2005}.

It has recently been observed that smaller galaxies, such as Leo I, can also harbor very massive black holes, as large as $3.3 \times 10^6\,M_{\odot}$ \citep{Majo_2021}, with a possible theoretical explanation given in the work of \cite{Pau_2014Sowing}. Additionally, the dynamical and kinematic properties of elliptical galaxies lead to the conclusion that some elliptical galaxies are the result of previous galaxy mergers or are still under a merging process \citep{Barnes_1992,Bekki_1998,Kormendy_2013}. The collision of galaxies results in a larger elliptical galaxy in which a binary system, formed by the central MBHs of each galaxy, perturbs the distribution of stars and compact objects. The binary exchanges energy with the surrounding objects generating a slingshot process among the surrounding stars \citep{Quinlan_1996,Sesana_2006,Khan_2018,Rasskazov_2019} that hardens the binary up to the point in which GWs lead the orbital evolution. Eventually, the MBHs merge, and some studies show that during this process, the interaction between the two MBHs can enhance the formation of extreme mass-ratio inspirals (EMRIs) as one of the MBHs scatters compact objects toward the other MBH \citep{Bode_2013,Naoz_2022,Mazzolari_2022}. Mass segregation is also present, bringing the heaviest objects to the central part and enhancing the number of compact objects that can potentially form inspiraling systems. Under this scenario, EMRIs are formed by a combination of different dynamical processes, such as relaxation, which is chaotic in nature but also the Kozai-Lidov mechanism \citep{Kozai_1962, Lidov_1962}.

We focus on Intermediate Mass-Ratio Inspirals (IMRIs) formed after a close encounter in which an unbound or loosely bound stellar-mass black hole (BH) is captured by an IMBH due to gravitational waves (GWs) emission. The formation of IMRIs in dense systems, such as globular clusters, has been studied by several authors \citep{PauMarc_2006,Konstantinidis_2013,Pau_2018IMRIs,ArcaSedda_2021}; here we explore the scenario in which IMRIs are formed as a result of close encounters in merging elliptical galaxies, where we assume that one of the interacting elliptical galaxies delivers an IMBH that travels towards the center of the gravitational potential due to dynamical friction; the IMBH moves in a dense system that contains a large number of compact objects, such as BHs, which facilitates close encounters. Interacting elliptical galaxies generate an ideal environment for IMRI formation as the merging process delivers an additional number of compact objects to an elliptical galaxy that already contains an old stellar population and, potentially, a large number of compact remnants. Furthermore, these systems can be important GW sources for space-borne detectors such as the Laser Interferometer Space Antenna (LISA) and TianQin as the peak frequency of the GWs emitted when the binary is formed is within the LISA \citep{Barack_2004,Pau_2007,LISA_2017} and TianQin \citep{TianQin0,TianQin} detection band. The Laser Interferometer Gravitational-Wave Observatory (LIGO) has already detected the GW signal of a binary black hole merger that resulted in a 150 M$_{\odot}$ black hole \citep{LIGO_2020}, which can be considered as an IMBH in the low-mass end; however, due to its detection range LIGO can not detect systems with masses $\gtrsim 10^3$ M$_{\odot}$ \citep{LIGO_O3_2020}; only space borne GWs detectors will be able to detect IMRIs containing IMBHs in the $10^3 - 10^5$ M$_{\odot}$ mass, providing essential information on the number and location of existing IMBH, the growth of MBHs, and the evolution of galaxies and their satellites. 

The IMRI formation mechanism is described in Section \ref{sec:gwcapture}, where we also show that this process is efficient for MBHs with masses $\lesssim 2\times 10^6$ M$_{\odot}$ making IMBHs ideal candidates to perform such captures. In Section \ref{sec:Binary_formation} we obtain the encounter rate that results in a binary system; however, some of these binaries do not become IMRIs as the system can be perturbed by encounters with other objects. For this reason, in subsection \ref{subsec:IMRIs} we compute the formation rate of the binaries that become successful IMRIs, i.e., systems in which the captured BH crosses the event horizon of the IMBH after inspiraling for a merger timescale $T_{\rm GW}$ without being perturbed by incoming objects. We also compute the peak frequency of the GWs emitted at the moment of formation to finally present our discussion and conclusions in Section \ref{sec:conclusions}.

\section{Binary formation by GW capture and formation rates}
\label{sec:gwcapture}

Inspiraling systems are formed when two-body relaxation processes perturb the orbit of an object $m_2$ already orbiting a massive black hole, such that after just one pericentre passage, the orbit evolves only due to GWs emission \citep{Hopman_2005,Pau_2007, Pau_2018, Pau_2020}. However, inspiraling binary systems can also be formed by objects in unbound or loosely bound orbits resulting from the interaction between two merging galaxies. In this environment, the high densities and the scattering of objects can enhance close encounters between objects, and if, during the encounter, enough energy is released as GWs, the encounter results in a new binary system that can become an inspiral. Therefore, not only would relaxation processes lead to inspiraling systems formation, but binaries formed by GW capture can also be an important source of inspiraling systems. We describe the formation process of a successful inspiraling system based on the following two conditions:
\begin{itemize}
\item[(i)] A single close encounter between two unbound or loosely bound, compact objects $M$ and m$_2$ dissipates enough energy in the form of GW to form a binary system.
\item[(ii)] The system merges in a timescale $T_{\rm GW}$ such that the orbit is not affected by encounters with surrounding objects.
\end{itemize}

The distance at which condition (i) is satisfied was derived by \citet{QuinlanShapiro1987} assuming that the orbit can be described as a parabolic orbit with a pericentre located at $r_{\rm p} \lesssim r_{\rm GW}$, where
\begin{align}
r_{\rm GW}=\left[\frac{85\sqrt{2}\pi G^{7/2} M m_{2}(M+ m_{2})^{3/2} }{12 c^5 v_{\rm rel}^2} \right]^{2/7} ,
\label{eq:r_perimax}
\end{align}
is the maximum passage distance, or capture radius, required to form a binary; $v_{\rm rel}$ is the relative velocity between the two objects, $G$ the gravitational constant, and $c$ the speed of light in vacuum. 

Equation~\eqref{eq:r_perimax} is found by equating the kinetic energy $\mu v_{\rm rel}^2/2$, where $\mu$ is the reduced mass of the system, to $\Delta E$, the amount of energy emitted as GW, derived from the time average of the energy emission rate \citep{Peters1964} (see also \citet{Turner_1977}, for an alternative formulation)
\begin{equation}
\left< \frac{dE}{dt}\right>= - \frac{32}{5} \frac{G^4 M^2 m_2^2 (M+m_2)}{c^5 a^5 (1-e^2)^{7/2}} \left( 1+\frac{73}{24} e^2 + \frac{37}{96}e^4 \right). 
\label{eq:DeDt}
\end{equation}
Integrating the energy emission $\left< dE/dt \right>$ over one orbital period $P=2 \pi (a^3/G(M+m))^{1/2}$, and setting the pericentre distance $r_{\rm p} =r_{\rm GW}= a(1-e)$, we obtain that for $e\to 1$, 
\begin{align}
|\Delta E| =& \frac{64 \pi }{5\times 8 \sqrt{2}}  \left( 1+\frac{73}{24} + \frac{37}{96} \right) \frac{G^{7/2} M^2 m_2^2  (M+m_2)^{1/2}}{c^5 r_{\rm p}^{7/2} }, \nonumber \\
&= \frac{85 \pi}{12\sqrt{2}} \frac{G^{7/2} M^2 m_2^2  (M+m_2)^{1/2}}{c^5 r_{\rm GW}^{7/2} }.
\label{eq:DeltaE}
\end{align}

\begin{figure}
\includegraphics[width=0.53\textwidth]{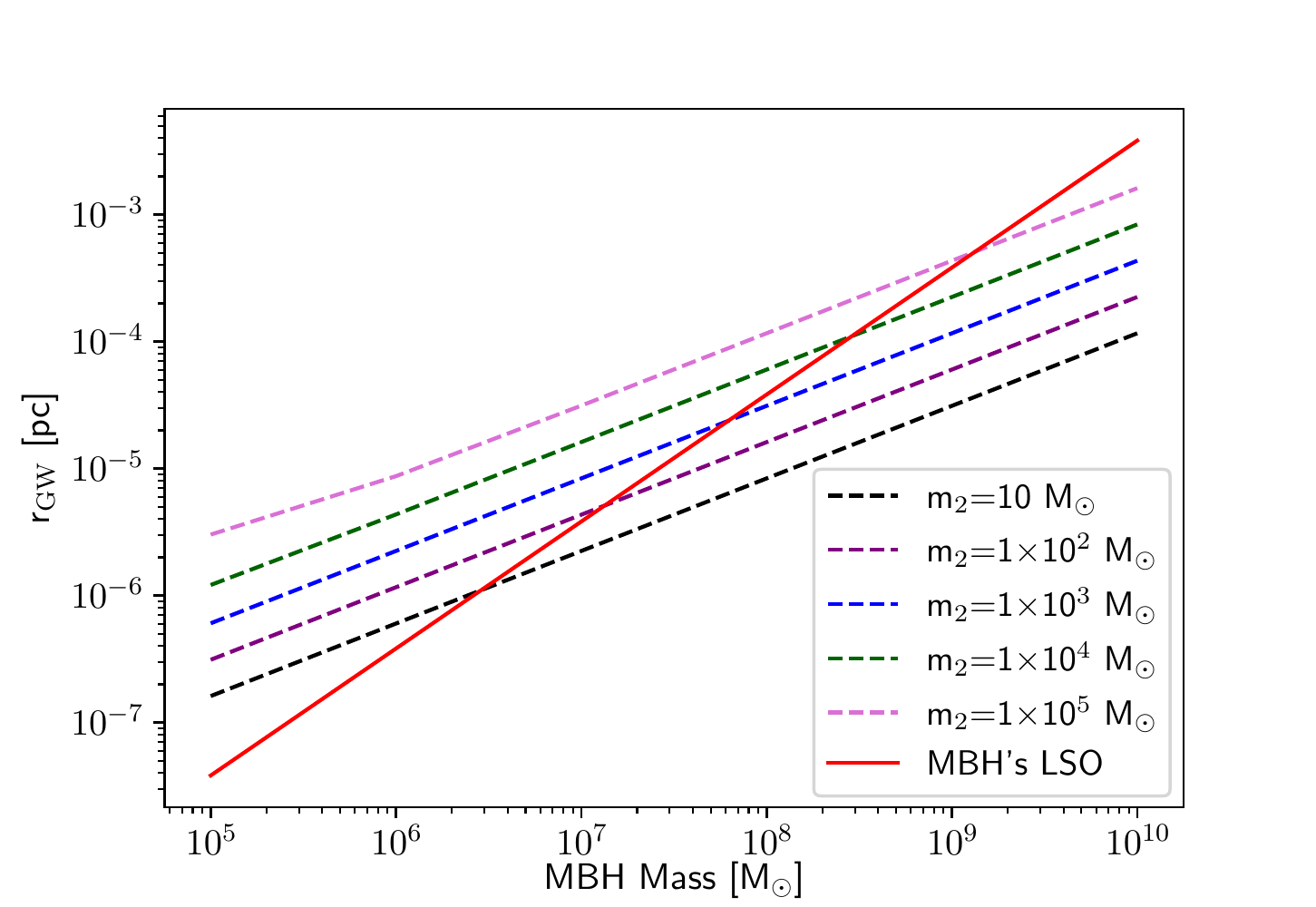}
\caption{GW capture radius $r_{\rm GW}$ as function of the massive black hole mass, $M\in$ [1$\times 10^5$, 1$\times 10^{10}$] M$_{\sun}$, for a set of black holes m$_2=[10,1\times 10^2, 1\times 10^3, 1\times 10^4, 1\times 10^5$] M$_{\sun}$. Capture by emission of GW occurs when $r_{\rm GW}$ lies above the red line which indicates the position of the last stable orbit computed for each MBH.}
\label{fig:rgwrplunge}
\end{figure}

We consider close encounters that result in binary systems with pericentre distances $r_{\rm p} = r_{\rm GW}$. The eccentricity and semimajor axis of these newly formed binaries is given by
\begin{align}
a_{\rm GW}=&\frac{GMm_{2}}{2|E_{\rm f}|}, \label{eq:sma_GW} \\
e_{\rm GW}=& 1-\frac{r_{\rm p}}{a_{\rm GW}},  \label{eq:e_GW}
\end{align}
where $E_{\rm f}$ is the final energy of the system
\begin{align}
E_{\rm f}= &\Delta E+ \mu v_{\rm rel}^2 / 2.
\end{align}

Close encounters can lead to direct plunges if $e_{\rm GW}\gtrsim e_{\rm plunge}$, as the orbit takes m$_2$ within the last stable orbit (LSO) of the MBH. The value of the maximum eccentricity $e_{\rm plunge}$ is given by
\begin{align}
e_{\rm plunge}=&1-\frac{r_{\rm LSO}}{a_{\rm GW}}, 
\label{eq:e_plunge}
\end{align}

where $r_{\rm LSO}= 4 r_{\rm S}$, and  $r_{\rm S}=2G M/c^2$ is the Schwarzschild radius. The distance $r_{\rm LSO}$ is derived from the critical angular momentum, $J_{\rm crit}$, that allows a non-relativistic particle in a highly eccentric orbit around a Schwarzschild black hole to avoid direct plunge \cite{ShapTeu_1983}
\begin{align}\noindent
J_{\rm crit} =\frac{4 G M}{c}. 
\end{align}
Any particle with $J < J_{\rm crit}$ plunges into the black hole; therefore, we assume that an orbit with $J= J_{\rm crit}$ has a pericentre distance located at the LSO, $r_{\rm p}=r_{\rm LSO}$. The total angular momentum is 
\begin{align}
&J =[GM a(1-e^2)]^{1/2}.
\end{align}
Taking $J= J_{\rm crit}$ and $e\to 1$, we obtain $r_{\rm p}=r_{\rm LSO} =4 r_{\rm S}$. 

Figure~\ref{fig:rgwrplunge} shows the position of the last stable orbit, $r_{\rm LSO}$, of a set of MBHs with masses $M\in$ [1$\times 10^5$, 1$\times 10^{10}$] M$_{\sun}$ and the capture radius $r_{\rm GW}$ computed for m$_2$=[10, 1$\times 10^2$, 1$\times 10^3$, 1$\times 10^4$, 1$\times 10^5$] M$_{\sun}$. If the mass of the MBH is $\gtrsim 2\times 10^9 $M$_{\sun}$, the capture radius $r_{\rm GW}$ lies inside the LSO for the m$_2$ objects, indicating that the incoming object crosses the LSO before losing a significant amount of energy by GW radiation and rapidly plunges into the SMBH. On the contrary, MBHs with M $\lesssim 2.8\times 10^6 $M$_{\sun}$ can capture all the considered m$_2$ objects. We particularly focus on intermediate-mass black holes with M$\in[1\times 10^3,1\times 10^5] $ M$_{\sun}$ because, as shown in subsection \ref{subsec:IMRIs}, for larger masses, condition (ii) is not satisfied and IMRIs are not formed.

\begin{figure}
\includegraphics[width=0.475\textwidth]{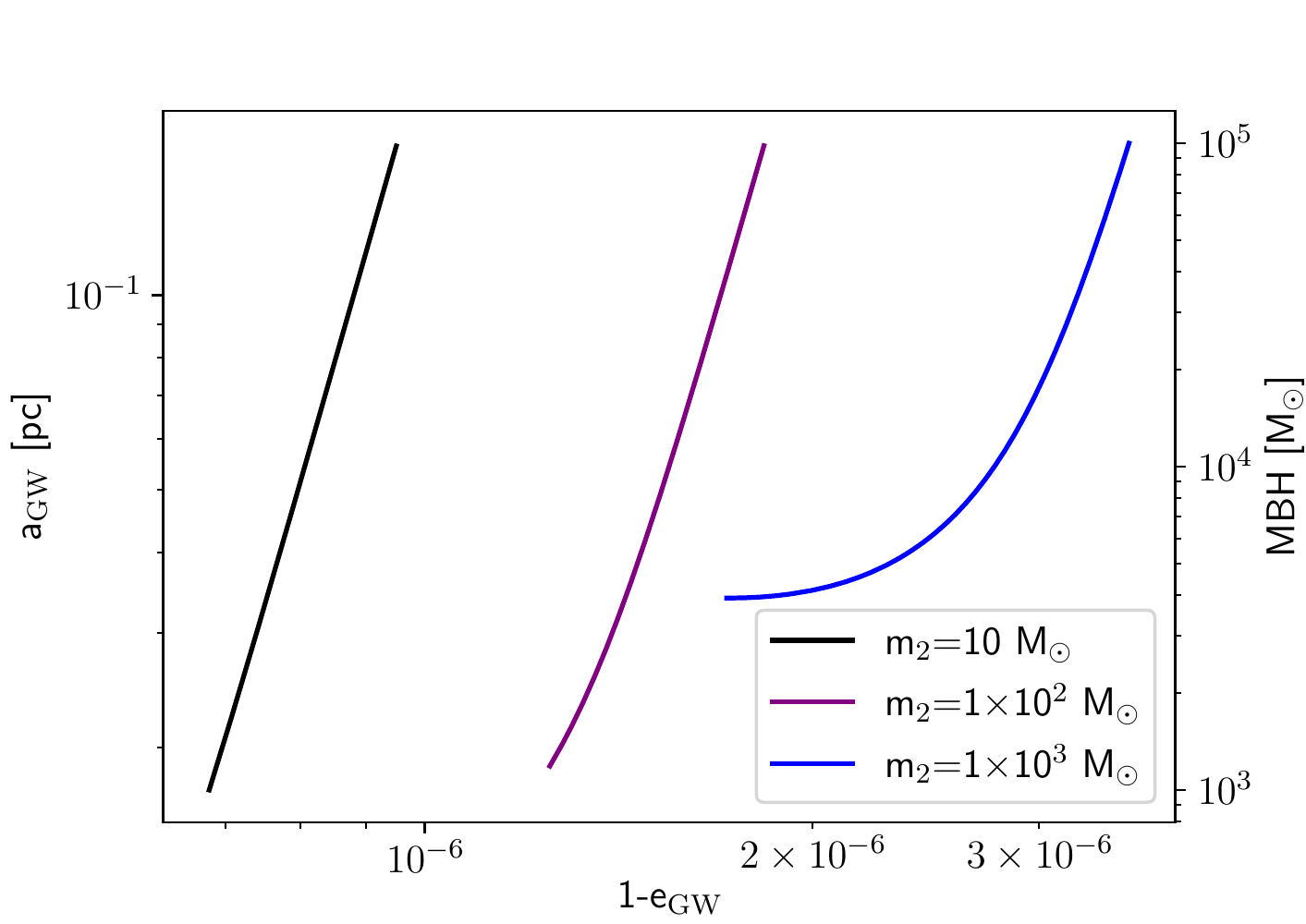}
\caption{Initial semimajor axis and eccentricity of binary systems formed by GW capture as a function of the MBH mass (right axis).}
\label{fig:smae}
\end{figure}

Figure~\ref{fig:smae} shows the initial semimajor axis and eccentricity of the binary systems formed by close encounters. These systems are initially formed with high eccentricities, $e_{\rm GW} \in (0.999996,0.999999)$, and initial semimajor axes ranging from $1\times10^{-2}$ pc to $1\times10^{-1}$ pc. Due to the high eccentricity of these systems, we can assume that the pericentre of the orbit remains approximately constant during the inspiral and compute the merger timescale as in \cite{Hopman_2005} by obtaining the time in which the initial energy $\epsilon_0 = GMm_2/2a$ grows to infinity
\begin{equation}
T_{\rm GW}= \int_{\epsilon_0}^{\infty} \frac{d\epsilon}{(d\epsilon / dt)} \approx \frac{2 \epsilon_0 P}{|\Delta E|},
\end{equation}
where $P=2\pi (a^3/G(M+m))^{1/2}$ is the orbital period and $|\Delta E|$ is given by Equation~\eqref{eq:DeltaE}, thus
\begin{align}
T_{\rm GW}=\frac{24 \sqrt{2}}{85} \frac{c^5 a^4}{G^3 M m_2 (M+m_2)} (1-e)^{7/2}.
\label{eq:mergertimescale}
\end{align}
IMRIs formed by GW capture merge in a timescale that ranges from $\sim 3\times 10^4$ yr to $\sim 8\times10^4$ yr,  $T_{\rm GW}$ is short yet important as there is a large number of interacting galaxies in the observational volume of the space-borne GW detectors.

\section{Binary formation rate}
\label{sec:Binary_formation}

The rate at which condition (i) is satisfied is estimated as the inverse of the timescale at which an encounter between $M$ and m$_2$ occurs at a distance $r_{\rm p}$. For a single compact object, the encounter timescale is obtained as \citep{BinneyTremaine}
\begin{align}
t_{\rm enc} \approx \frac{1}{nSv_{\rm rel}}, 
\label{eq:t_enc}
\end{align}
where $n$ is the number density of compact objects around the MBH within a radius R, and $S$ is the capture cross section (see Equation \ref{eq:capturecrosssection} below). 

Numerical simulations indicate that the evolution of the central velocity dispersion during a galaxy merger can be separated into three phases: oscillation, phase mixing, and dynamical equilibrium. During these phases, the velocity dispersion value remains above the 70$\%$ and up to the 200$\%$ of its value at equilibrium, but a random measurement of the velocity would likely fall near the equilibrium value \citep{Stickley_2012, Stickley_2014}. For this reason, we assume that the M-$\sigma$ relation holds \citep{Tremaineetal2002}
\begin{align}
\sigma \sim 200 \, \left( \frac{M }{ 10^8 M_{\sun} } \right) ^{1/4} \text{km s}^{-1},
\label{eq:Msigma}
\end{align}
and we take $v_{\rm rel} \approx \sigma$. Assuming that the relative velocity between the objects remains between $0.7 \sigma$ and $2 \sigma$, it is important to include gravitational focusing, so we estimate the capture cross section as
\begin{equation}
S= \pi r_{\rm p}^2 \left(1 + 2 \frac{G(M+m_{2})}{\sigma^2r_{\rm p}}\right). 
\label{eq:capturecrosssection}
\end{equation}
Therefore, the formation rate of a single binary can be written as
\begin{align}
\Gamma_{\rm GW} & = n\, S \sigma \nonumber \label{eq:encounterrate}\\
& =  \pi\,  n  \left[ 2 \widehat{M} \, r_{\rm p}^2 + G  \widehat{M}^{-1} (M+m_2) r_{\rm p} \right], \\
\widehat{M} &=\left(\frac{M}{M_{\odot}}\right)^{1/4}.  \nonumber
\end{align}
The estimate given by Equation~\eqref{eq:encounterrate} provides the rate at which a close encounter generates a burst of GWs that dissipates enough energy to form a binary system; yet, a fraction of the newly formed binaries will be disrupted by forthcoming encounters. Condition (ii) must be satisfied to form an inspiraling system that will eventually merge, i.e., the merger timescale $T_{\rm GW}$ (Equation \ref{eq:mergertimescale}) must be shorter than the encounter timescale $t_{\rm enc}$ (Equation \ref{eq:t_enc}). 

The formation of inspiraling systems by GW capture is favored in systems with high density, where close encounters between MBHs and BHs in unbound or loosely bound orbits occur. Merging elliptical galaxies are systems that can provide such features; however, their specific characteristics are not well known, so we assume typical values for a galactic center to estimate the formation rate of inspiraling binary systems with $e_{\rm GW}\lesssim e_{\rm plunge}$ such that $T_{\rm GW} < t_{\rm enc}$. We find that inspirals are formed between BHs and MBHs within the IMRI mass range, as for larger MBHs, the merger timescale is longer than the encounter timescale, and the binary can be disrupted by future encounters.

\subsection{Intermediate mass-ratio inspirals}
\label{subsec:IMRIs}

\begin{figure}
\includegraphics[width=0.5\textwidth]{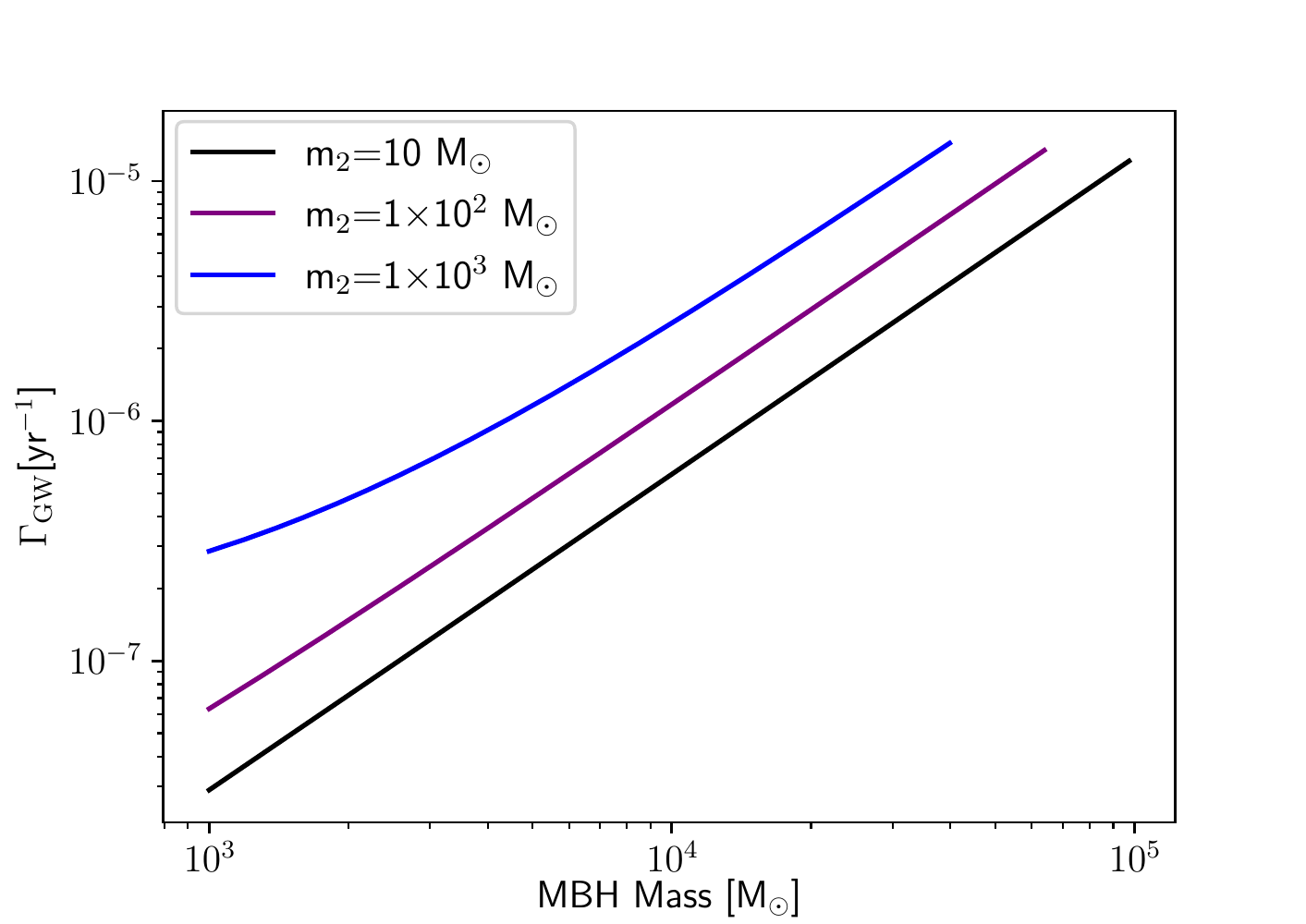}
\includegraphics[width=0.5\textwidth]{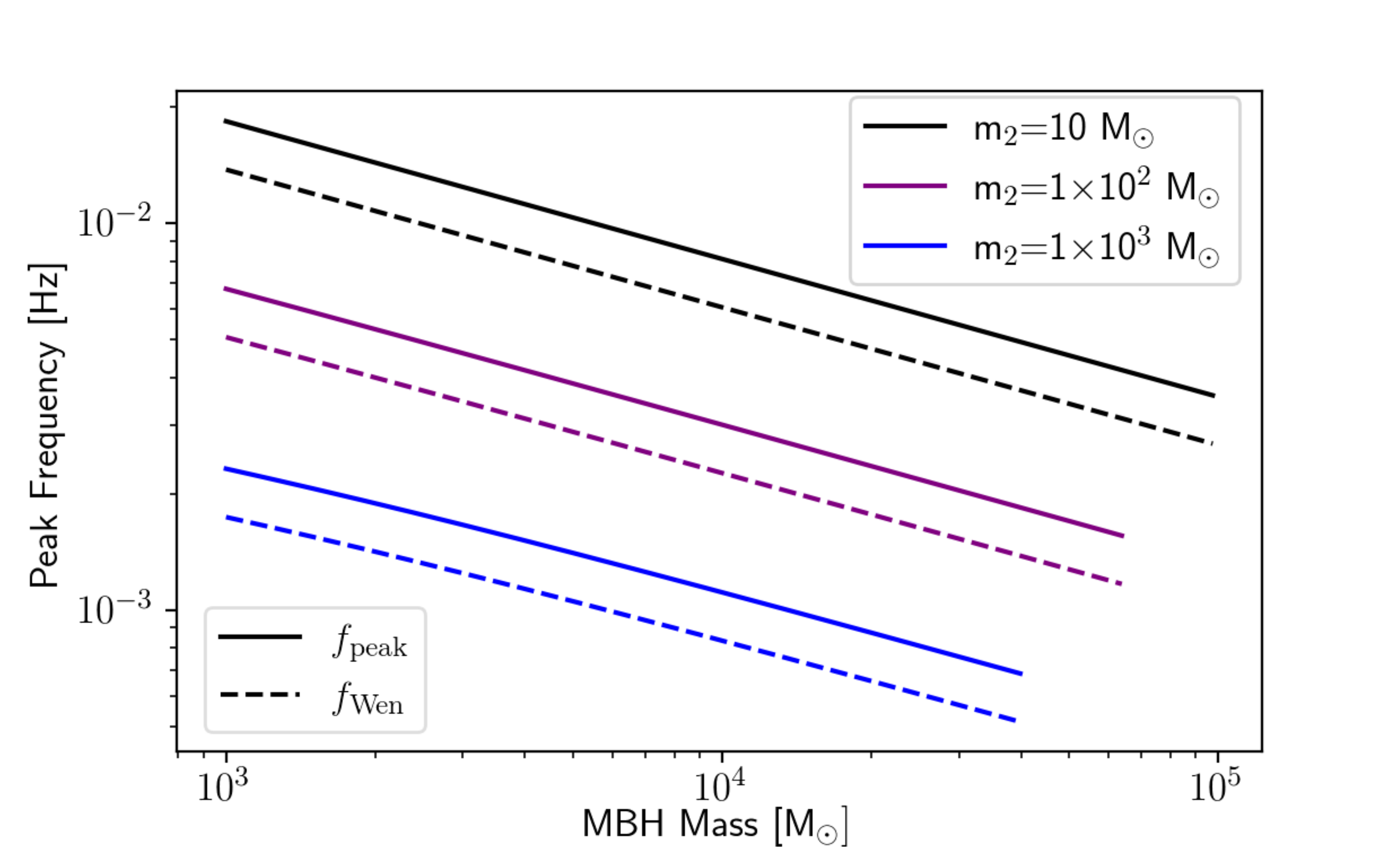}
\caption{Upper panel: Formation rate $\Gamma_{\rm GW}$ of a single IMRI as a function of the massive black hole mass, with $n=1\times 10^6$ pc$^{-3}$.
Lower panel: Peak frequencies $f_{\rm peak}$ (solid line), and $f_{\rm Wen}$ (dashed line) at the moment of binary formation as a function of the MBH mass.}
\label{fig:Nrate_peakfreq}
\end{figure} 

 The characteristics of the environment created by merging elliptical galaxies are not well known. In the central parts of elliptical galaxies, the surface brightness can be described either by a core where the brightness flattens and remains approximately constant, or by a cusp where the brightness rises towards the centre following a power law. During a galaxy merger, the mass ratio between the interacting galaxies can determine the final distribution of the central parts; nevertheless due to mass segregation, the heavier objects will travel to the central parts of the system. Is specially in this region, where dynamical interactions can enhance close encounters that favor the GW capture. For a more accurate description one needs to estimate the velocity dispersion $\sigma$, the number density $n$, and the number of objects $N$ within a given radius $R$. This is part of an on-going study and will be published elsewhere. However, in this section, and as a first-order approximation to be improved, we assume typical values for a galactic center to estimate the formation rates. By taking $n=1\times 10^6$ pc$^{-3}$ as a typical numerical density, we find that inspirals, i.e., systems with $e_{\rm GW}\lesssim e_{\rm plunge}$ and $T_{\rm GW} \lesssim t_{\rm enc}$, are formed when the mass of the MBH ranges from 10$^3$ M$_{\odot}$ to $\lesssim 10^5$ M$_{\odot}$, and $m_2\in$(10, 10$^3$) M$_{\odot}$. Due to the mass ratio, these systems are considered IMRIs, and their formation rate, obtained with Equation~\eqref{eq:encounterrate}, ranges from $\Gamma_{\rm GW} \sim 10^{-8}- 10^{-5}$yr$^{-1}$ as shown in the upper panel of Figure~\ref{fig:Nrate_peakfreq}.

With equations~\eqref{eq:sma_GW} and~\eqref{eq:e_GW} we obtain the orbital parameters a$_{\rm GW}$ and $1-e_{\rm GW}$, shown in Figure~\ref{fig:smae}, of the binaries that become IMRIs. The initial semimajor axis of these systems ranges from $1\times10^{-2}$ pc to $1\times10^{-1}$ pc, and as the systems are highly eccentric, the GW power is radiated predominantly through the $\bar{n}$=2.16 (1-$e$)$^{-3/2}$ harmonic; the peak frequency emitted by the IMRI is given by \citep{Farmer_2003} 
\begin{align}
f_{\rm peak}= &\frac{ 2.16 }{ P}\, (1-e)^{-3/2}, \label{eq:fpeak} \\
P=&2\pi a^{3/2}[G(M+m_{2})]^{-1/2}. \nonumber
\end{align}

\cite{Wen_2003} finds a fit to the emitted GW power and derives the following peak frequency
\begin{equation}
f_{\rm Wen}= \frac{ \sqrt{ G (M+m_2)} }{ \pi}\, (1+e)^{1.1954} \frac{1}{a(1-e^2)^{1.5}}. 
\label{eq:fwen}
\end{equation}

Frequencies obtained with \eqref{eq:fpeak} and \eqref{eq:fwen} differ by a constant factor $f_{\rm Wen}/f_{\rm peak}\sim 0.75$, remaining within the detection band of the space-based GW detectors. LISA will be able to observe frequencies of $\sim 10^{-4}$ to $10^{-1}$ Hz, although its sensitivity band is centered at $f \sim 3 \times 10^{-3}$ Hz \citep{Barack_2004}, while the TianQin detector can detect frequencies ranging from $\sim 10^{-4}$ to 1 Hz. The lower panel of Figure~\ref{fig:Nrate_peakfreq} shows $f_{\rm Wen}$ and $f_{\rm peak}$ for different IMRIs; the minimum frequency $\sim 10^{-4}$ Hz corresponds to a system composed by IMBHs of $10^5$ and $10^3$ M$_{\odot}$, while the maximum frequency $\sim 10^{-2}$ Hz, to a system composed by black holes of $10^3$ M$_{\odot}$ and $10$ M$_{\odot}$.

Considering a simplified scenario in which during a galaxy merger, only the central velocity changes and oscillates between the 70$\%$ and the 200$\%$ of its value at equilibrium $\sigma$ \citep{Stickley_2012}, we obtain that the semimajor axis of the binaries formed through GW capture, $a_{\rm GW}$, changes by a factor that ranges between $\sim 2$ and $\sim 0.2$, while the formation rate, $\Gamma_{\rm GW}$, varies from $\sim 1.7~ \Gamma_{\rm GW}$ to $\sim 0.3~ \Gamma_{\rm GW}$. Nevertheless, the peak frequencies shown in Figure~\ref{fig:Nrate_peakfreq} change by a factor $\sim 0.7$ when the central velocity reduces to 0.7$\sigma$, and by a factor $\sim 1.8$ when the central velocity goes up to 2$\sigma$, remaining within the LISA and TianQin detection band. 

Capture by GW emission can also form binary systems composed of BHs and SMBHs with a mass M $\lesssim 2.8 \times 10^6$ M$_{\odot}$. However, the encounter timescale is shorter than the merger timescale, and $\Gamma_{\rm GW}$ indicates how often a binary is formed only to be disrupted after a timescale $t_{\rm enc}$. In this scenario, the specific characteristics of the environment are of great importance, if $t_{\rm enc}$ is short, these sources are not likely to be detected as they will be rapidly disrupted; on the contrary, if $t_{\rm enc}$ is long enough, we could detect these binaries while they inspiral by a timescale $t_{\rm enc}$. In the case of $M=5\times 10^5$M$_{\odot}$, $t_{\rm enc}\sim 8\times10^4$ yr, which could still be important as the event rates are relatively high $\Gamma_{\rm GW} \sim 10^{-4}$ yr$^{-1}$.

IMRIs have been studied in globular clusters (GCs) \citep{PauMarc_2006,ArcaSedda_2021} and in some cases, the resulting IMBH can leave the system if the kick velocity acquired during the merger is higher than the escape velocity of the GC \citep{Holley_2008,Konstantinidis_2013}. The magnitude of the kick velocity depends on the mass ratio and  the spins of the merging black holes \citep{Lotsuo_2008}, but in a merging galaxy scenario, the large escape velocities would allow the IMBH to stay in the host merging galaxy, and the inspiral formation process can continue. Detecting and locating an IMRI will give us valuable information on the existence and evolution of the IMBH population.

The event rates in the upper panel of Figure \ref{fig:Nrate_peakfreq} are obtained for a single galaxy merger system; to compute the number of IMRIs forming in the local universe in merging galaxies, we need to know the number of merging galaxies; there are several galaxy merger observations, but the number of merging galaxies in the local universe is highly uncertain. Estimates based on observational data indicate that about the $1-4\%$ of all the galaxies between redshifts $z = 0.005$ and $z = 0.1$ are merging galaxies \citep{GalaxyZoo_2010a,GalaxyZoo_2010b,Man_2016, Pearson_2020}, but this fraction is not constant. \cite{Lotz_2008} finds that the observed fraction of merging galaxies varies from $\sim 44\%$ at $z\sim 0.3$ to $\sim 21\%$ at $z\sim 1.1$,  which is relevant as the LISA detection volume for IMRIs goes up to, at least, redshift $z\sim 1$ \citep{LISA_2017}. We now estimate the IMRI formation rate in merging galaxies per comoving volume as a function of the redshift from the volume averaged number density of ongoing galaxy merger events per unit time for $z < 1.5$, described in \cite{Lotz_2011},
\begin{equation}
\Gamma_{\rm gal}= \frac{f_{\rm m} n_{\rm gal}}{T_{\rm obs}} \approx 2.6 \times 10^{-3} (1+z)^{0.4}\text{Gyr}^{-1} \text{Mpc}^{-3}, 
\label{eq:gamma_gal}
\end{equation}
where $f_{\rm m}$ is the fraction of merging galaxies obtained from a galaxy sample, $T_{\rm obs}\sim 0.2$ Gyr is an average time during which the merger can be observed,  and $n_{\rm gal}$ is the comoving number density of galaxies, which we consider constant, $n_{\rm gal}\sim 10^{-3}$ Mpc$^{-3}$. By multiplying $\Gamma_{\rm gal}$ from Equation \eqref{eq:gamma_gal} by $T_{\rm obs}$ we can obtain the number of merging galaxies at a given redshift per Mpc$^{-3}$, and assuming that for each merging systems, the IMRI formation rate is given by $\Gamma_{\rm GW}$, the volume averaged IMRI formation rate in merging galaxies as a function of the redshift is 
\begin{equation}
\overline{\Gamma}_{\rm IMRI}\approx 5.2 \times 10^{5} (1+z)^{0.4} \times \Gamma_{\rm GW}~ \text{Mpc}^{-3} \text{yr}^{-1}
\end{equation}

\begin{figure}
\includegraphics[width=0.5\textwidth]{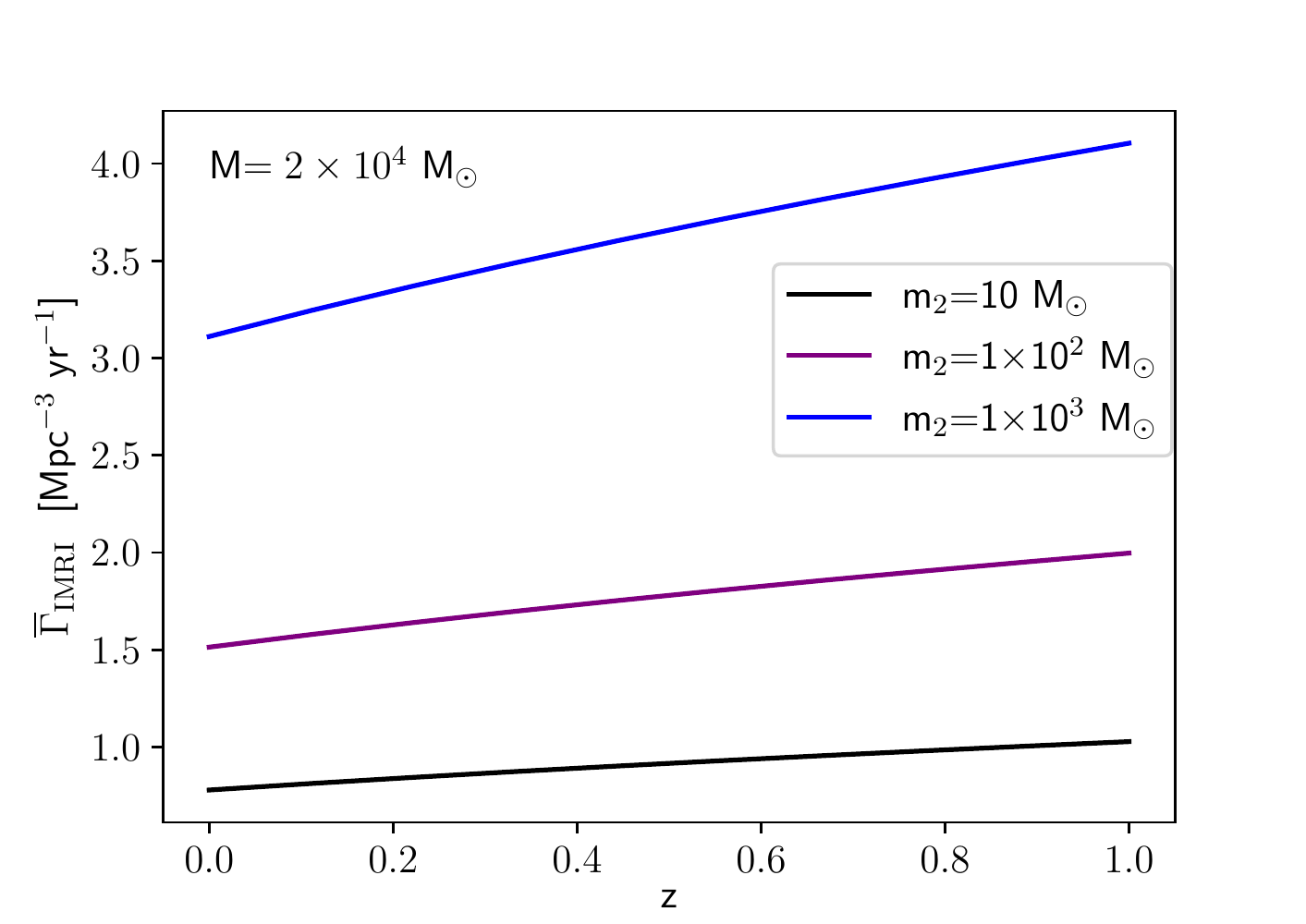}
\caption{IMRI formation rate per Mpc$^{-3}$ in merging galaxies as a function of the redshift for M=$1\times 10^4$ M$_{\odot}$ and m$_2$=[10,$1\times 10^2$,$1\times 10^3$] M$_{\odot}$.}
\label{fig:gal_rate}
\end{figure} 
In Figure \ref{fig:gal_rate}, we show $\overline{\Gamma}_{\rm IMRI}$ up to redshift $z=1$, for an IMBH of $2\times 10^4$ M$_{\odot}$ and m$_2$=[10,$1\times 10^2$,$1\times 10^3$] M$_{\odot}$. These event rates, added to the event rates estimated for IMRIs in GC could form a GW background, similarly to EMRIs \citep{Bonetti_2020}, if the GW signals do not reach a sufficiently high signal-to-noise ratio (SNR) due to the orbital parameters or the distance to the source. Note that $\overline{\Gamma}_{\rm IMRI}$ is averaged over Mpc$^{-3}$, and that this represents an upper limit as it accounts for all types of merging galaxies. GW capture can also occur in non elliptical merging galaxies, but the presence of large amounts of gas, star formation processes, and the dynamical structures in non elliptical galaxies require a more detailed description. 

\section{Discussion and conclusions}
\label{sec:conclusions}

Due to their old stellar population, the interaction between merging elliptical galaxies can provide BHs with a wide range of masses, creating a favorable environment for inspiraling systems formation. We study binary formation due to the emission of GWs in close encounters that result in intermediate mass-ratio inspirals considering a dense system generated by two merging elliptical galaxies, in which at least one of them brings an IMBH into the central part of the system. We find that inspirals are efficiently formed if the mass of the MBH is between 10$^3$ M$_{\odot}$ and $\lesssim$10$^5$ M$_{\odot}$, and $m_2\in$(10, 10$^3$) M$_{\odot}$. Also, space-borne detectors such as LISA and TianQin will be able to detect such systems at formation and different stages of evolution. Larger MBHs with masses $\lesssim 2.8\times 10^6$ M$_{\odot}$ can also capture BHs by GW emission in close encounters; nevertheless, these systems do not merge as new encounters perturb the orbit before the binary merges.

We note that this process can also occur in single galaxies or globular clusters; however, the old population and lack of gas in elliptical galaxies, the large densities, and the dynamical processes of two merging galaxies can enhance close encounters and the chances of having at least one IMBH in the system. Nonetheless, this scenario and the evolution of IMRIs need to be explored in more detail. As a first-order approximation, we take a typical galactic-center density to obtain the formation rate of IMRIs formed by GW capture and find that it is about one order of magnitude lower than the event rates obtained for relaxation processes \citep{Hopman_2005,Pau_2007,VVA_2022} e.g., for M$\lesssim 10^5$M$_{\odot}$, the rate for relaxation processes is $\sim 10^{-4}$yr$^{-1}$, while $\Gamma_{\rm GW} \sim 10^{-5}$yr$^{-1}$. However, inspirals formed by relaxation processes assume that the IMBH is fixed at the central part of a density cusp, but due to the mass range of the IMBHs and the galaxy-merging environment, a moving IMBH can not be ruled out. A moving IMBH affects the relative velocity between the objects (see Equation \ref{eq:capturecrosssection}), but the binary formation process by GW capture can continue. Additionally, the formation of IMRIs in a galaxy merger is interesting as after the IMRI merges, the resulting IMBH can be kept in the galaxy and continue with the capture process. 

A deeper analysis of a GW background formed by IMRIs at different evolution stages formed not only in merging galaxies but also in globular clusters is needed. Binaries that do not become IMRIs can also be sources of GWs by a timescale given by $t_{\rm enc}$; if this timescale is long enough, these systems could increase the number of sources that we expect to detect within the detection band of LISA and TianQin. IMRIs formed by GW capture have a merger timescale of $\sim 10^4 - 10^5$ yr; however, the volume averaged IMRI formation rate in merging galaxies, $\overline{\Gamma}_{\rm IMRI}$, ranges from 0.7 Mpc$^{-3}$ yr$^{-1}$ to 4 Mpc$^{-3}$ yr$^{-1}$ in the case of a IMBH of $2\times10^4$ M$_{\odot}$, and m$_2$=[10, $1\times10^2$, $1\times10^3$] M$_{\odot}$, which indicates an important amount of IMRIs in the observable volume of the space-born GW detectors. 

\section*{Acknowledgments}
We are indebted with Anna Lisa Varri and Marc Freitag for many conversations.
This research is funded by the Science Committee of the Ministry of Education and Science of the Republic of Kazakhstan (Grant No. AP09259383).
VVA acknowledges support from CAS-TWAS President's Ph.D. Fellowship Program of the Chinese Academy
of Sciences \& The World Academy of Sciences. PAS'
work has been supported by the National Key R\&D Program of China
(2016YFA0400702), the National Science Foundation of China (11721303)
and the 111 Project under Grant No. B20063.

\section*{Data Availability Statement}
The data underlying this article will be shared on reasonable request to the corresponding author.

\bigskip
\bibliographystyle{mnras}           %
\bibliography{ellipt.bib}    %
\label{LastPage}
\end{document}